\documentclass[runningheads]{llncs}
\usepackage{hyperref}
\usepackage{balance}
\usepackage{graphicx}
\usepackage[ruled,vlined]{algorithm2e}
\usepackage{tikz}
\usetikzlibrary{shapes,arrows,calc,positioning}
\usetikzlibrary{automata}
\usepackage{amsmath,amssymb,mathtools}
\usepackage{xcolor}
\usepackage{url}
\usepackage{enumitem}
\usepackage{subcaption}
\usepackage{psfrag}
\usepackage{epstopdf}
\usepackage{pgfplots}
\usepackage{gensymb}
\usepackage{psfrag}
\usepackage{epstopdf}
\usepackage{tablefootnote}
\usepackage{caption}
\usepackage{ifthen}
\usepackage{cleveref}
\pgfplotsset{compat=newest}
\usetikzlibrary{plotmarks}
\usepackage{multirow}
\usepackage[belowskip=-0pt,aboveskip=0pt]{caption}
\setlist[itemize]{noitemsep, topsep=0pt}

\begin{document}
\title{Analyzing Occupancy-Driven Thermal Dynamics in Smart Buildings}
\titlerunning{Occupancy-Driven Thermal Dynamics in Smart Buildings}
%
\author{Khaza Anuarul Hoque\inst{1} \and
Nathalie Cauchi\inst{2} \and
Alessandro Abate\inst{2}}
\authorrunning{K.A.Hoque et al.}
%
\institute{University of Missouri, Columbia MO, USA \and
University of Oxford, Oxford, UK\\
\email{\{khaza.hoque, nathalie.cauchi, aabate\}@cs.ox.ac.uk}}
\maketitle              
\begin{abstract}
The fact that a proper HVAC control strategy can reduce the energy consumption of a building by up to 45\% has driven significant research in demand-based HVAC control. This paper presents a novel framework for modeling and analysis of thermal dynamics in smart buildings that incorporates building's thermal properties, a stochastic occupancy model and heating strategies. Each zone of a building is modeled with the help of discrete time Markov rewards formalism where the states represent the occupancy of that zone (either occupied or empty), and the state rewards incorporate the thermal dynamics and heating strategy. To demonstrate the applicability of our proposed framework,  we evaluate and compare six different heating strategies for the two zone scenario of a university building. The obtained quantitative results from the PRISM probabilistic model checker show that one of the evaluated control strategies (viz. selective strategy) satisfies our requirement in terms of maintaining the occupants' comfort while being up to 13.5 times more cost effective when compared to the other evaluated strategies. Such evaluations demonstrate the framework's ability to assist in selecting the control strategy tailored around the occupancy pattern and building's thermal property. 

\end{abstract}
\section{Introduction}
The Internet-of-things (IoT) has enabled the realization of \textit{Smart Buildings}, which aim to deliver useful building services that are efficient, cost effective, reliable, an ubiquitous ensuring occupants' comfort. Smart building automation systems aim to reduce energy consumption of buildings by operating the Heating, Ventilation and Air Conditioning (HVAC) system in a more efficient way, thereby to addressing energy and environmental concerns \cite{esterly20142013}. The first step to achieve optimal performance is to construct a quantitative modeling framework of the building itself  from energy consumption perspective. This is a difficult task due to the inherent complexity of all the different thermal interactions in the rooms, together with all the unmeasured disturbances as a result of changes in weather or occupancy~\cite{Ma2012:IEEE_Mag}. Furthermore, the cost of energy adds more complexity to the problem. For instance, in the winter season, given a building topology, occupancy pattern and other factors (e.g., solar heat gain), what should be an effective heating strategy while considering the heating cost and occupants' comfort? An intelligent heating strategy can be to predict the expected temperature of a zone ahead of time based on the expected occupancy and underlying thermal model of the zone, and turn the heating on/off accordingly. This also allows to schedule heating when energy prices are lower, hence reducing the overall building operating cost. Heating a room when it is empty is of course a counter-intuitive choice in many cases, however, if preheating a room is cheaper (based on the electricity cost) and capable of keeping the room warm for next few hours, then suddenly such a strategy can become a preferred option. To assess such strategies, a novel building thermal model and analysis framework is thus required, which motivates our work in this paper.

\noindent \emph{Related works.} Different modelling techniques for building energy management have been proposed in the literature ~\cite{Miller2013:IEEE,Nagy2015:IEEE,walker2017application}, %
the most popular of which is to use a Resistance Capacitance (RC) network to describe the underlying system process dynamics of buildings and zones ~\cite{Rel:braun2001evaluating,Rel:xu2008simplified}. 
In~\cite{peva}, a seventh-order RC network is used to model two zones followed by a model order reduction technique based on the underlying
physical quantities. Using this order reduction methodology, a library of thermal models with varying orders of complexity has further been presented. 
In~\cite{gouda2000low}, the authors proposed a second-order RC-network for modeling the heat conduction through walls, while~\cite{Deng10Thermal} adopted the RC-network framework and then proposed a technique for aggregation-based model reduction for large buildings. The heat gained within a zone in a building is highly dependent on the occupancy of the zone. Consequently, 
several advanced models have been proposed in the literature to model building occupancy \cite{rel33,chen2015modeling}. In~\cite{EbadatBottegalVaragnoloEtAl2015}, proposed a two-tier deconvolutional approach is proposed to model occupancy profile of a room by using pilot data to identify models for the CO$_2$ and temperature from which occupancy is estimated. The occupancy estimation problem is modeled in~\cite{Jiang2016132} as a regression framework. Alternative techniques based on Markov principles have been presented in~\cite{Ryu20161,McKenna201530}. The authors in~\cite{chen2015modeling} make use of inhomogenous Markov chains to detect the presence of occupants where the states represented the increment of occupancy in a zone. In the work~\cite{dong2011building}, Gaussian mixture models were first used to identify the sensor which yielded the highest information gain for detecting the number of occupants in the zones. The selected features are then fed into a hidden Markov model to estimate the number of occupants. Since occupancy also stimulates the demand of other heat sources, such as lighting or electronic devices, occupancy has become an important factor in estimating building energy consumption \cite{rel11,rel22}. To summarize, researchers from academia and industry has made substantial efforts to in model building thermal dynamics and occupancy, but they have mostly dealt with them as two separate problems instead of an unified model that captures both the thermal dynamics and \textit{stochastic} occupancy pattern. \\

\noindent \emph{Contributions of this paper.} In this paper, we aim to bridge this gap and propose a novel framework, and develop an unconventional model for analyzing the temperature evolution of zones in a building. The proposed model incorporates the thermal dynamics, stochastic occupancy pattern, and heating sources (e.g., radiators) in a zone. More specifically, the thermal dynamics of each zone is modeled using the discrete time-inhomogeneous Markov reward model (DTIMRM) formalism from the given building topology as RC equations \footnote{An RC network can be generated from the building information models (BIM), or by using the state-of-the-art tools such as EnergyPlus~\cite{crawley2000energy}}, occupancy pattern of a zone, and the radiator status from the given heating strategy. The DTIMRM \emph{states} represent the occupancy pattern (calculated from a given dataset) of the zone (occupied or empty) and the thermal dynamics of the zone including the given heating strategy is captured as \emph{state rewards}. The time-inhomogeneity of the model is then resolved by unfolding it over a specified number of time steps, thus converting an DTIMRM in a discrete time-homogeneous Markov reward model (DTMRM). Afterwards, we perform \emph{parallel composition} of these DTMRMs, each of which represent a zone in a building. 

The composed model is then evaluated to quantify the expected temperature of each zone with respect to different heating control strategies. The heating strategies are preset based on the predefined thermal conditions. This falls under the framework of rule based control which is commonly employed in the heating of buildings~\cite{ALIMOHAMMADISAGVAND2018167}. Performing a heating strategy analysis allows us to highlight the performance of the developed framework under different settings, while demonstrating the ability of the framework to guide us in selecting the correct control strategy depending on the specific needs. Using our proposed framework, computation of the probabilistic reachability does not require a large state-space, as witnessed when using probabilistic reachability analysis tools such as FAUST$^2$~\cite{soudjani2015faust} which requires gridding over the individual zone temperature range. Such quantitative analysis can help to adopt the most effective and cost efficient heating strategy that maintains a specified range of temperature (for occupants' comfort) during the intended duration of operation as shown in our obtained results.  A key advantage of our proposed model is its capability to deal with the time-inhomogeneity, that none of the above mentioned existing approaches can handle. The framework is also generic enough and can be easily extended to model additional number of zones. As mentioned earlier, the building topology is modeled as RC equations (treated as inputs) which are converted to Markov models later. Hence, different building typologies can be evaluated using our approach with minimal effort. This is also true for evaluating different occupancy patterns and heating strategies. 

For automated analysis of the proposed models, we use the Probabilistic Model Checking (PMC) technique \cite{PMC}, and more specifically, the PRISM model checker \cite{prism}. PMC is mainly based on the construction and analysis of a probabilistic model, typically a Markov chain or a Markov process. These models are constructed in an exhaustive fashion. In discrete-event simulations, approximate results are generated by averaging results from a large number of random samples. In contrast, PMC applies numerical computations to provide accurate results which makes our approach more reliable. Although the approach in \cite{Deng10Thermal} also used a Markov chain for modeling thermal dynamics of a building, where temperature of zones is represented as states, such a model cannot incorporate thermal gains from internal or external sources. In other words, it can only model the conduction between the zones via the surface between them (e.g., walls). Our proposed framework overcomes this limitation by using the Markov reward model formalism. In addition, due to the use of formal verification tools, the results obtained using our framework can aid in certifying \emph{\textbf{safety-critical building facilities}}, such as hospitals, medical labs, military and defense-related infrastructures, nuclear power plants, food processing plants and so on.\\

\noindent \emph{Paper organization.} The outline of the paper is as follows. Section 2 presents some background about probabilistic model checking. In Section~\ref{sec:BTM}, we describe the framework in detail while Section~\ref{Sec:Res} outlines the obtained results. In Section~\ref{Sec:discuss}, we discuss some important aspects of our proposed framework approach, and finally, Section~\ref{Sec:Conc} concludes the work with directions of future research.


\section{Preliminaries}
\subsection{Probabilistic model checking}
\label{subsec:Prelim:PMC}
Model checking~\cite{clarke1986automatic} is a well-established technique used in both industry and academia to verify the correctness of finite-state concurrent models. \emph{Probabilistic Model Checking (PMC)} \cite{PMC} deals with systems that exhibit stochastic behavior, and is based on the construction and analysis of a probabilistic model of the system, typically a Markov chain. We make use of the DTMRM \cite{DTMRM} formalism based on a discrete time Markov chain (DTMC) for modelling the occupancy and thermal dynamics. 
\begin{definition}[Discrete time Markov chain (DTMC)] \label{def:DTMC}
	Let AP be a fixed, finite set of atomic propositions. A (labelled) DTMC $\mathcal{D}$ is a tuple $(S,s_{init},TL,P,L)$ where S is a finite set of states, $s_{init} \in S$ is the initial state, $TL$ is a set of transition labels denoting the transitions between states, $P: S \times S \rightarrow [0,1]$ is a probability matrix such that $\sum_{s' \in S} p(s,s') =1$ for all $s \in S$, and $L: S \rightarrow 2^{AP}$ is a labelling function that assigns to each state $s \in S$ the set $L(s)$ of atomic propositions that are valid in $s$. The \textit{time-inhomogeneous} version of DTMC $\mathcal{D}$ (denoted as $\mathcal{D}_{ih}$) has the probability matrix as a function of time, e.g., $P^k: S \times S \rightarrow [0,1]$ given $k \in \mathbb{N}$. 
\end{definition}

\begin{definition}[Discrete time Markov reward model (DTMRM)] \label{def:DTMRM}
	A discrete-time reward model (DTMRM) $\mathcal{M}$ is a pair $\mathcal{(D, \rho)}$ where DTMC $\mathcal{D} = (S,s_{init},TL,P,L)$ and $\rho : S \rightarrow \mathbb{R}_{\ge 0}$ is a reward assignment function that associates a real reward to any state in $S$. A Real number $\rho(s)$ denotes the state reward $s$. If DTMC $\mathcal{D}$ in the DTMRM pair is a time-inhomogeneous DTMC ($\mathcal{D}_{ih}$), then such a model is considered as a discrete time-inhomogeneous reward model (DTIMRM).
\end{definition}

\begin{figure*}[h!]
	\centering
	{\includegraphics[width=\textwidth]{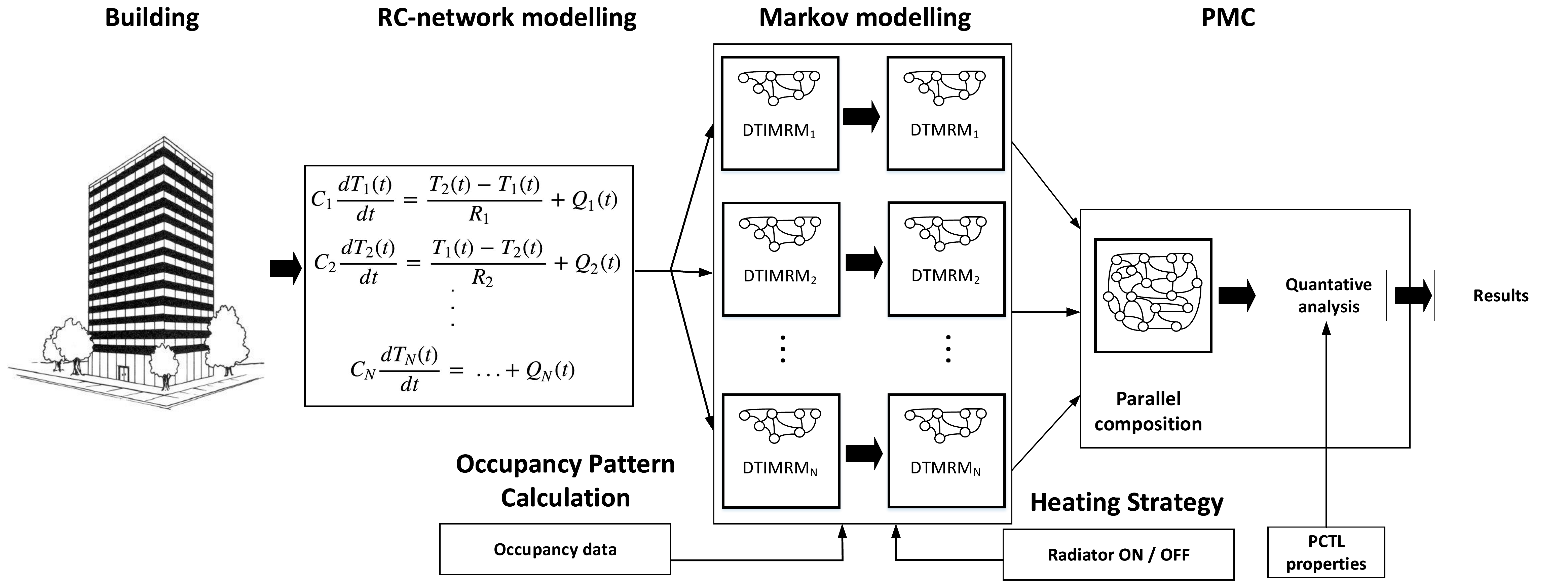}}
	\caption{Proposed modelling and evaluation framework.}
	\label{fig:method}
	\vspace{-3mm}
	\end{figure*}
	
\noindent \emph{Path and cumulative rewards.} Let $\mathcal{M}$ be a DTMRM with underlying DTMC $\mathcal{D} = (S,s_{init},TL,P,L)$ and reward function $\rho$. A finite path $\sigma$ with length $l$ is a sequence $s_0 \rightarrow \dots \rightarrow s_{l-1}$ with $s_i \in S$ and $p(s_i,s_{i+1}) > 0$ for $0 \leq i < l-1$. Let $Pr_s$ denote the unique probability measure as defined in \cite{pathDef} on the set of paths that start in state $s$. The cumulative reward along a finite path $\sigma$ with length $l$ is defined as $\rho(\sigma) = \sum_{i \geq 0}^{l-1} \rho(s_i)$. 
In the Probabilistic model checking approach using DTMCs or DTMRMs, the properties are usually expressed in some form of temporal logic such as Probabilistic Computation Tree Logic (PCTL) \cite{PCTL}, a stochastic variant of the well-known Computation Tree Logic (CTL) \cite{CTL}. 

\subsection{PRISM modeling and property specification} 
PRISM~\cite{prism} is a well known probabilistic model checking tool for the formal modeling and verification of stochastic systems. The current version of the tool supports four types of probabilistic models: discrete-time Markov chains (DTMCs), continuous-time Markov chains (CTMCs), discrete-time Markov decision processes (MDPs) and probabilistic timed automata (PTA). Models in PRISM are specified using a high-level modelling language based on the Reactive Modules formalism~\cite{alur1999reactive}.



To analyze a PRISM model, it is required to specify one or more properties. Since PRISM is a probabilistic model checker, the property specification language in PRISM is based on the temporal logic and includes operators from PCTL, CSL~\cite{baier1999approximative} and its extensions\cite{haverkort2002model}. An example of a PCTL property in PRISM with its natural language translation is: $ \mathcal{\textbf{P}}_{\geq 0.95} [\textbf{F} ~complete]$ - ``The probability of the system eventually completing its execution successfully is at least 0.95". Note, the \textbf{F} operator in PRISM property specification language refers to the \textit{eventually}~($\Diamond$) operator in PCTL~\cite{PMC}. Additional properties can be specified by adding the notion of rewards. Each state (and/or transition) of the model is assigned a real-valued reward, allowing queries such as: $\mathcal{\textbf{R}}_{=?} [\textbf{F} ~success] $ - ``What is the expected reward accumulated before the system successfully terminates?" The \textbf{R} operator in PRISM property specification language extends the PCTL with the notion of rewards~\cite{PMC}. Rewards can be used to specify a wide range of measures of interest, for example, the total operational costs and the total percentage of time the system is available. In this paper, we utilize the notion of \emph{cumulative rewards} (using the $\mathcal{\textbf{C}}$ operator) to evaluate the temperature of a zone in a building. A detail description of the PRISM language, property specification and probabilistic model checking using the PRISM tool can be found in \cite{PMC}.

\section{Thermal model of a Building}\label{sec:BTM}
A high-level description of the proposed modeling framework is depicted in Figure~\ref{fig:method}. Given a building and its corresponding topology, we first construct the equivalent \textit{RC-network model} made up of $N$ zones. Note, the zones should be connected to each other through walls. This RC-network model is converted into a state-space models which allows for ease of conversion into a \textit{Markov model}. We also perform an \textit{occupancy pattern calculation} which computes the probability of occupancy in each zone during each time step. For each zone, we construct an equivalent DTIMRM. Afterwards, we unfold them to a specified number of time steps which converts them into equivalent DTMRMs. These models incorporate the calculated \textit{occupancy pattern} and the \textit{heating strategy} to evaluate. We further perform the \textit{parallel composition }of these DTMRMs to obtain the whole building Markov model. Quantitative analysis is then performed on this model through the use of PMC technique (using a probabilistic model checker). 
For this paper, we consider two teaching rooms (zones) within the Department of Computer Science, University of Oxford and restrict ourselves to having two heat sources in each zone: (i) radiators \footnote{Use of radiators for heating rooms are still common in many European universities and offices, including the University of Oxford. However, this does not restrict the proposed framework to be applied to the forced-air heating systems, which is very common in the US. This will be further clarified in Remark~\ref{rmk:forced-air}.} and (ii) occupants. The teaching rooms are connected back to back through walls. The rooms are used between 8 am and 5 pm during term time. The stochastic occupancy pattern of these zones are calculated from our developed dataset containing the past occupancy record of a zone. A layout of these rooms is shown in Figure \ref{fig:Layout}(a) and Figure \ref{fig:Layout}(b).

\subsection{RC-network representation}\label{subs:RC}
A building thermal model is constructed by combining parameters of thermal interaction between two zones separated by a solid surface (e.g., walls, windows, ceilings, and floors). For the sake of simplicity, here we ignore the inter-zone convective heat transfer that occurs through the open doors and hallways. A lumped parameter model of combined heat flow across a surface is modeled as a simple RC-network, with current and voltage being analogous to heat flow and temperature, respectively. In this modeling framework, the capacitance are used to model the total thermal capacity of the wall, and the resistances are used to represent the total resistance that the wall offers to the flow of heat from one side to the other. 
	\begin{figure}[h]
   \centering      
   \includegraphics[width=0.95\textwidth]{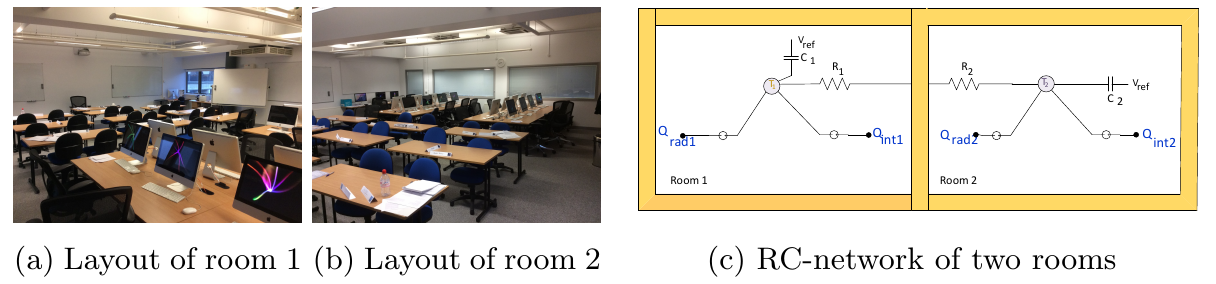}     
  \caption{Layout of teaching rooms and their RC-network representation.}\label{fig:Layout}
\end{figure}


A building topology can be represented as a directed graph $\mathcal{G} = (\mathcal{V}, \mathcal{E})$, where $\mathcal{V}:=\{1, . . . , N\}$ denotes the set of $N$ nodes of the graph. A node represents a physical zone (e.g., a room or a hallway). Each node $i \in \mathcal{V}$ corresponds to a temperature $T_i$ and has an associated capacitance $C_i$. The set $\mathcal{E} \subset \mathcal{V} \times \mathcal{V}$ denotes the set of all edges which represent pathways for conductive heat transports. More specifically, each edge $(i, j) \in \mathcal{E}$ represents the conductive thermal interaction between the nodes $i$ and $j$ represented by a resistance $R_{i} \in \mathbb{R}_{+}$. This is a different approach compared to \cite{deng2014structure}, where the building topology was represented using an undirected graph and the authors considered a point on the surface (such as the wall) as a physical zone, whereas we ignore that. The thermal dynamics of a multi-zone building, described by a directed graph $\mathcal{G}$, is modeled by a system of first order coupled linear differential equations as follows, for $i = 1, . . . , N$:

\begin{equation}
C_i \frac{dT_i(t)}{dt} = \sum_{j \in \mathcal{N}_i} \frac{T_j(t) - T_i(t)}{R_{i}} + {Q}_i(t) \label{eqn:thermal}
\end{equation}

\noindent where $\mathcal{N}_i:= \{j \in \mathcal{V} : j \ne i,(i, j) \in \mathcal{E}\}$ denotes the set of neighbors connecting to the node $i$. The heat gain term ${Q}_i(t)$ is the rate of thermal energy entering the node $i$ from all sources, such as occupants, lighting etc., and radiators other than the conduction from neighboring nodes. For instance, let us consider two rooms connected back to back to one another and are depicted using Figure \ref{fig:Layout}(c) following~\cite{maasoumy2011model}, where $V_{ref}$ is a reference temperature. The thermal model is composed of two zones having an average zone temperature $T_1,\;T_2$, which exchange heat with one another assuming that the heat is only gained in a zone via its occupants and radiators. To build the equivalent RC network the  following assumptions have been applied:
\begin{enumerate}
	\item Similar thermal properties between heat zones.
	\item Uniform temperature across the walls.
	\item Negligible solar gain and heat exchange with outside through walls. (Note that this is not a limitation of the proposed modeling approach. How to incorporate these factors in the proposed model has been addressed in Section 5.)  
	\item The air in each zone having one temperature across its volume.
\end{enumerate}

Using Equation \eqref{eqn:thermal} we obtain the equations for each zone:
\begin{align}
C_1 \frac{dT_1(t)}{dt} &= \frac{T_2(t) - T_1(t)}{R_{1}} + {Q}_1(t)  \notag \\
C_2 \frac{dT_2(t)}{dt} &= \frac{T_1(t) - T_2(t)}{R_{2}} + {Q}_2(t) \label{eqn:Zone}
\end{align}
where, ${i \in \{1,2\}}$ represents the respective zone and ${Q}_i(t) = {Q}_{rad_i}(t) + {Q}_{int_i}(t)$, in which stochastic terms ${Q}_{int_i}(t)$ and ${Q}_{rad_i}(t)$ represent the internal heat gain (i.e., the rate of heat generated by the occupants, equipment, lights, etc.) and heat gains from the radiators respectively, and can be specified as follows:

\[{Q}_{int_i}(t)= 
\begin{cases}
{Q}_{int_i}, & \text{if~the~room~is~occupied}\\
0, & \text{otherwise}
\end{cases}
\]

\[{Q}_{rad_i}(t)= 
\begin{cases}
{Q}_{rad_i}, & \text{if~the~radiator~is~on}\\
0, & \text{otherwise}
\end{cases}
\]
 
Equation \eqref{eqn:Zone} can be rewrittten in the form of a linear time invariant state space model as follows:
\begin{equation} \label{eqn:AT}
\dot{T} = \hat{A}T + \hat{B}U 
\end{equation}
where the state vector ${T} := [{T}_1, . . . , {T}_{N}]^T$ denotes the temperature of each zone and $U$ denotes the heat gain vector ${Q}:= [{Q}_1, . . ., {Q}_N]^T$. Both the transition rate matrices $\hat{A}$ and $\hat{B}$ have dimensions of $N \times N$. The entries of $\hat{A}$ matrix are given by:
\[\hat{A}_{ij}= 
\begin{cases}
0, & \text{if~} j \ne i, (i,j) \notin \mathcal{E}\\
\frac {1}{(C_iR_{i})}, & \text{if~}  j \ne i, (i,j) \in \mathcal{E}\\
- \sum_{k \ne i} \hat{A}_{i}, & \text{if~}  j = i, (i,j) \in \mathcal{E}
\end{cases}
\]

and the entries of matrix $\hat{B}$ can be expressed as:
\[\hat{B}_{ij}= 
\begin{cases}
\frac {1}{(C_i)}, & \text{if~}  i = j, (i,j) \in \mathcal{E}\\
0, & \text{otherwise} 
\end{cases}
\]

For instance, the example shown in Figure \ref{fig:Layout} (c) with two zones, can be expressed as:
\[
\begin{split}
   \left[ \begin{array}{cc}
\dot{T}_1\\
 \dot{T}_2
\end{array} \right] &= \left[ \begin{array}{cc}
\frac{-1}{R_1 C_1} & \frac{1}{R_1 C_1} \\
\frac{1}{R_2 C_2} & \frac{-1}{R_2 C_2}
\end{array} \right]\left[ \begin{array}{cc}
T_1 \\
T_2
\end{array} \right] + \left[ \begin{array}{cc}
\frac{1}{C_1} & 0\\
0 & \frac{1}{C_2}\\
\end{array} \right]  \left[ \begin{array}{cc}
{Q_1}\\
{Q_2}
\end{array} \right]\\
&= \left[ \begin{array}{cc}
\frac{-1}{R_1 C_1} & \frac{1}{R_1 C_1} \\
\frac{1}{R_2 C_2} & \frac{-1}{R_2 C_2}
\end{array} \right]\left[ \begin{array}{cc}
T_1 \\
T_2
\end{array} \right] + \left[ \begin{array}{cc}
\frac{Q_1}{C_1}\\
\frac{Q_2}{C_2}\\
\end{array} \right]\\
\end{split}
\]  

Based on the data obtained from \cite{peva} ($C_1 = 1.3700, C_2= 1.000, R_1=1.7429, R_2= 5.5897$), we discretize the model using the forward Euler method \cite{euler} with a sampling time of 1 hour ($\Delta = 1~hr$), and obtain the A and B matrices (discretized version of $\hat{A}$ ~and~ $\hat{B}$ matrix, respectively)

\[
A = 
\begin{bmatrix}
0.7001 & 0.2999\\
0.3007 & 0.6993
\end{bmatrix}, ~
B = 
\begin{bmatrix}
0.7299&0 \\
0&1
\end{bmatrix} \]

\begin{remark} \label{rmk:thermalDyn}
	We are interested in modeling a university building with classrooms (zones), where the duration of each class is around 1 hour. Moreover, as mentioned earlier, the thermal gain of a zone is assumed only from the occupants or from the radiator. Since thermal dynamics of zones are relatively slow with time constants in tens of minutes \cite{gouda2000low} to hours \cite{tashtoush2005dynamic}, the dynamics of the radiators are replaced by static gains in this paper without significant loss of accuracy. This also justifies the time interval of one hour while discretizing from continuous-time to discrete-time. 
\end{remark}

\subsection{Markov modeling of occupancy and thermal dynamics}


In this section, we show how the thermal dynamics of zones can be encoded using Markov chains. For such modeling, we use the discretized A and B matrices as in the previous section. The equation for the expected temperature of zones at time step 1 can be written as :

\begin{equation}\label{eqn:T1}
T[1] = AT[0] + {Q[0]}
\end{equation}


\noindent where, ${Q}[0]$ represents the combined gain from the occupants and radiator from the previous time step (time step 0, e.g. initial values in this case). Similarly, the expected temperature for the zones in $k$-th time step can be derived as:
\begin{equation}\label{eqn:Tn}
T[k] = \underbrace{A^{k}T[0]}_\text{step 0} + \underbrace{\sum_{m=1, n=2}^{m=k-1, n=k} A^m{Q}[k-n]}_\text{steps from 1 to k-1} + \underbrace{{Q}[{k-1}]}_\text{step k}
\end{equation}

\begin{figure*}[t!]

	\centering
	\resizebox{0.9\textwidth}{!}{
	{\begin{tikzpicture}[font=\sffamily]

\tikzset{node style/.style={state, 
		minimum width=1cm,
		line width=0.5mm,
		fill=white!20!white}}

\node[node style, initial above, fill = black!30!green] at (0, 0)  (s0)     {$s_0^0$};
\node[node style, fill = yellow!30] at (2, -2)  (s1)     {$s_1^1$};
\node[node style, fill = black!30!green] at (2,  2)  (s2)     {$s_2^1$};
\node[node style, fill = black!30!green] at (6, 2)   (s4) {$s_4^2$};
\node[node style, fill = yellow!30] at (6,-2)   (s3) {$s_3^2$};
\node[node style, fill = black!30!green] at (11, 2)  (sN) {$s_{2n}^K$};
\node[node style, fill = yellow!30] at (11,-2)  (sN-1) {$s_{2n-1}^K$};
\node[]           at (9, 2) (dots1) {\Large  $\dots\dots$};
\node[]           at (9,-2) (dots2) {\Large $\dots\dots$};
\node[node style, fill = lightgray] at (13,0)  (sN+1) {$s_{2n+1}^{K+1}$}; 
\draw[every loop,auto=right,line width=0.5mm,>=latex,draw=black,fill=black]
(s0) edge[bend right=20, auto=right] node  [above, yshift=5pt] {$t_1$} node [below, xshift = -5pt] {$p_{vf}^{1}$} (s1) 
(s0) edge[bend left=20,  auto=right]  node [above, yshift=5pt] {$t_1$} node [below,yshift=-5pt, xshift = 5pt] {$p_{vv}^{1}$} (s2) 
(s2) edge[right=20, auto=right] node  [above] {$t_2$} node [below] {$p_{vv}^{2}$} (s4) 
(s2) edge[bend right=20, auto=right] node  [above, yshift=6pt] {$t_2$} node [below, yshift=22pt,xshift = -25pt] {$p_{vf}^{2}$} (s3) 
(s1) edge[bend right=20, auto=right] node  [above, yshift=6pt] {$t_2$} node [below, yshift=22pt, xshift = 28pt] {$p_{fv}^{2}$} (s4) 
(s1) edge[right=20, auto=right] node  [above] {$t_2$} node [below] {$p_{ff}^{2}$} (s3)
(s4) edge[right=10, auto=right] node  [above] {$t_3$} node [below] {$p_{vv}^{3}$} (dots1) 
(s4) edge[bend right=10, auto=right] node  [above, yshift=6pt] {$t_3$} node [below, yshift=5pt,xshift = -14pt] {$p_{vf}^{3}$} (dots2) 
(s3) edge[right=10, auto=right] node  [above] {$t_3$} node [below] {$p_{ff}^{3}$} (dots2) 
(s3) edge[bend right=10, auto=right] node  [above, yshift=15pt] {$t_3$} node [below, yshift=24pt,xshift = 27pt] {$p_{fv}^{3}$} (dots1) 
(dots1) edge[right=10, auto=right] node  [above] {$t_K$} node [below] {$p_{vv}^{k}$} (sN) 
(dots1) edge[bend right=10, auto=right] node  [above, yshift=18pt, xshift = -2pt] {$t_K$} node [below, yshift=5pt,xshift = -9pt] {$p_{vf}^{k}$} (sN-1) 
(dots2) edge[right=10, auto=right] node  [above] {$t_K$} node [below] {$p_{ff}^{k}$} (sN-1) 
(dots2) edge[bend right=10, auto=right] node  [above, yshift=15pt] {$t_K$} node [below, yshift=24pt,xshift = 27pt] {$p_{fv}^{k}$} (sN)


(sN-1) edge   [bend right=10, auto=right]  node [above, xshift=-7pt]{$t_{K+1}$} node [below]{1} (sN+1) 
(sN)   edge   [bend left=10, auto=right]   node [above, xshift= 5pt]{$t_{K+1}$} node [below]{1} (sN+1);  

\end{tikzpicture}}
	}
	\caption{DTMC of a zone for $K+1$ time steps.}
	\label{fig:room}
	\vspace{-4mm}
\end{figure*}
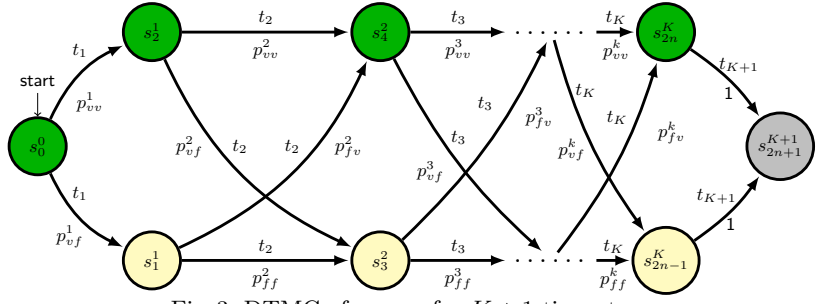

Since occupancy is a stochastic term, e.g., a zone can be either occupied or empty at a given time step, the Markov chain analogy is now clear. Before proceeding further, it is worth mentioning the assumptions of the proposed model as follows:

\begin{itemize}
	\item A zone can be either occupied or empty. If occupied, it is assumed to be fully occupied, otherwise empty.
	\item Since we are modeling classrooms, each time step (class duration) is assumed to be equal to 1 hour.
	\item The internal gain (per hour) of a zone from the occupants and zone's radiator is assumed to be constant. 
	\item Each zone has one radiator for heating purpose.
\end{itemize}

\begin{remark}\label{rmk:forced-air}
In contrast to the use of radiators for heating, the forced-air heating system is more popular in the US. In a forced-air system, the heating rate is modeled as a continuous variable, however, it can be abstracted to a single number. In order to do that, one needs to discretize the equation $\Delta H = C_{pa}\dot{m}_i^{in}(t)(T^s-T_i(t))$, where $C_{pa}$ is the specific heat capacitance of the supplied air at constant pressure, $T^s$ is the supply
air temperature (usually constant for a variable air volume system \cite{deng2013model}), and $\dot{m}_i^{in}$ denotes the mass flow rate of the supply air entering the $i$-th zone.
\end{remark}

\begin{figure}[!h]
	\centering
	\resizebox{0.4\textwidth}{!}{{\begin{tikzpicture}[font=\sffamily]

\tikzset{node style/.style={state, 
		minimum width=1cm,
		line width=0.5mm,
		fill=white!20!white}}

\node[node style, initial left, fill = black!30!green] at (0, 0)  (s0)     {$s_1$};
\node[node style, fill = yellow!30] at (4, 0)  (s1)     {$s_2$};

\draw[every loop,auto=right,line width=0.5mm,>=latex,draw=black,fill=black]


(s0) edge[bend right=20, auto=right]  node [below, xshift = -5pt] {$P_{vf}^k$} (s1)
(s1) edge[bend right=20, auto=right]  node [below, xshift = -5pt] {$P_{fv}^k$} (s0)

(s1) edge [loop above]  node {$P_{ff}^k$} (s1)
(s0) edge [loop above] node {$P_{vv}^k$} (s0);

\end{tikzpicture}}}
	\caption{DTIMRM of a zone.}
	\label{fig:DTIMRM}
\end{figure}
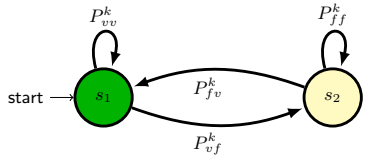

Given that a building $\mathcal{B}$ has $N$ zones, each zone can be modeled as a separate discrete-time inhomogeneous Markov reward models (DTIMRM) as a compact representation. In the DTIMRM model, the $states$ represent the occupancy of a zone (either occupied or empty) and we use the notion of $state~rewards$ to encode the thermal dynamics in the proposed Markov model based on Equation \eqref{eqn:Tn}. The rewards are collected at each time step based on the current state, and hence the expected temperature at $k$-th time step can be calculated as cumulative rewards up to time step $k$. The DTIMRM of a zone is shown in Figure \ref{fig:DTIMRM}. For clarity, the figure is color coded - the dark green node represents an empty zones and light yellow colored node represents an occupied zone. We assume that the zone was initially empty encoded by state 0.

\begin{definition}(zone-DTIMRM) The occupancy and temperature evaluation of a zone can be defined as discrete-time inhomogeneous Markov reward model (DTIMRM) $\mathcal{R}_{ih}$ which is a pair $(\mathcal{D}_{ih}, \rho^k)$ with time-inhomogeneous DTMC $\mathcal{D}_{ih} = (S, s_{init}, L, P^k)$ and a reward function $\rho^k$, where:
	
	\begin{itemize}
		\item $S = \{s_1, s_2\}$, represents if a zone is either occupied or empty;
		\item $ s_{init} = \{s_1\}$ is the initial state;
		\item $L: S \rightarrow 2^{AP}$, a labeling function that assigns each state $s_i \in S$ a label from the set of atomic propositions $AP = \{occupied, empty, heating_{on}, heating_{off}\}$;
		\item $P^k: S \times S \rightarrow [0,1]$ is the time dependent stochastic (occupancy) matrix that associates each pair $(s,s') \in S \times S$ the transition probability where $\sum_{s' \in S} p^k(s,s') = 1$ for all $s \in S$. More specifically: $P^k =    \begin{bmatrix}
		p_{ff}^k & p_{fv}^k \\    p_{vf}^k & p_{vv}^k \\
		\end{bmatrix}     
		$
		where $f$ and $v$ stands for \emph{occupied} and \emph{empty} states respectively, for instance, $p_{vf}^k = p^k(v,f)$.
		\item $\rho^k:S \rightarrow \mathbb{R}_{\ge 0}$, a time dependent reward function that assigns a reward to each state to encode the thermal dynamics and gains (from heating and occupant) to the states at a corresponding time step $k$ (explained in detail in Definition \ref{zone-DTMRM}). 
	\end{itemize}
\end{definition}

To resolve the time-inhomogeneity, we expand the model by unrolling the DTIMRM up to $K+1$ steps ($k \in K$), where $K$ denotes the time window (operating time) of the zones. Thus the DTIMRM is converted into a discrete-time Markov reward model (DTMRM). Figure \ref{fig:room} shows the Markov model of a zone for $K+1$ steps. From each empty state, there are two possible transitions labelled as $t_k$ with probability $P_{vv}^{k}$ and $P_{vf}^{k}$, which respectively represent the probability of being empty or occupied in the next time step (how we derive these transition probability is explained in the next subsection). Similarly, from each occupied state, there are two possible transitions labelled as $t_k$ with probability $P_{fv}^{k}$ and $P_{vv}^{k}$ that respectively denote the probability of being empty or occupied in the next time step. The grey colored node ($K+1$-th state) denotes the sink state. Note that, we use the same labels for labelling transitions at a specific time step, however only one of them will be triggered at a given time step. Formally, a zone can be redefined as follows:

\begin{definition}(zone-DTMRM)\label{zone-DTMRM} The occupancy and temperature evolution of a zone can be defined as discrete-time Markov reward model (DTMRM) $\mathcal{R}$ which is a pair $(\mathcal{D}, \rho)$ with time-homogeneous DTMC $\mathcal{D} = (S, s_{init}, TL, L, P)$ and a reward function $\rho: S \rightarrow \mathbb{R}_{\ge 0}$ that assigns a reward to each state $s_i^k \in S$, where:
	\begin{itemize}
		\item $S = \{s_{0}^0, s_{1}^1, s_{2}^1,s_{3}^2, s_{4}^2, ...., s_{2n}^K, s_{2n+1}^{K+1},\}$, where $s_{i}^k \in S$ represents $i^{th}$ state in $k^{th}$ time step and $k = {1,2, ... ,K+1}$;
		\item $ s_{init} = \{s_{0}^0\}$ is the initial state;
		\item $ TL = \{t_1, t_2, ... , t_{K+1} \}$ is a set of transition labels denoting the time steps and used for synchronization (explained in Definition \ref{parr-comp});
		\item $L: S \rightarrow 2^{AP}$ is a labeling function that assigns each state $s_i^k \in S$ a label from the set of atomic propositions $AP = \{occupied, empty, heating_{on}, heating_{off}\}$;
		\item $P: S \times S \rightarrow [0,1]$ is the stochastic (occupancy) matrix that associates each pair $(s,s') \in S \times S$ the transition probability where $\sum_{s' \in S} p(s,s') = 1$ for all $s \in S$. Since, in our model each odd numbered states denote an occupied zone and each even numbered states denote an empty state, for each time step $k$ the entries of the stochastic matrix $P = (p_{ij}^k)$ are given by:
		\[p_{ij}^k= 
		\begin{cases}
		p_{vf}^k, & \text{if~} i=even, j=odd, j-i = 1\\
		p_{vv}^k, & \text{if~} i=even, j=even, j-i = 2\\
		p_{ff}^k, & \text{if~} i=odd,  j=odd, j-i = 2\\
		p_{fv}^k, & \text{if~} i=odd,  j=even, j-i = 3\\
		1,        & \text{if~} i \geq 2n-1 \text{~and~} j=2n+1\\
		0, & \text{otherwise}\\
		\end{cases}
		\]

		
		
		
		
		
		\item $\rho$ is a reward function defined as follows: 
		\begin{equation} \label{eqn:reward}
		\rho(s_{i}^k) = 
		\begin{cases}
		\textbf{A}^{\theta} T[0], &\text{if~} s_{i}^k \in s_{init} \\
		\textbf{A}^{\theta-k} \mathcal{Q}^k_{i}, & \text{if~} \theta,K>k>0\\
		J^k_{i}, & \text{if } \theta = K\\
		0, &\text{otherwise}
		
		\end{cases}
		\end{equation}
		
		where $\textbf{A}^{\theta} = [X_{mj}, X_{m(j+1)}, ..., X_{mN}]$ is $1\times N$ matrix for the $m$-th zone given that $X = A^{\theta}$. The term $\theta$ defines the specific time of interest to evaluate the model (explained in Example 1) and the discretized $A$ matrix is derived from thermal dynamics as described in the previous subsection. For instance, if a building has two zones, then $\textbf{A}^{\theta}$ can be defined as $\textbf{A}^{\theta} = [X_{11}, X_{12}]$ for zone 1 ($m=1$) and $\textbf{A}^{\theta} = [X_{21}, X_{22}]$ for zone 2 ($m=2$). $T[0]$ is a $N \times 1$ matrix specified as $T[0] = [T_1[0], T_2[0], ..., T_N[0]]^T$, where $T_m[0]$ represents the initial temperature of $m$-th zones. $\mathcal{Q}^k_{i}$ is also a $N \times 1$ matrix specified as $\mathcal{Q}^k_{i} = [\mathcal{C}^k_{1i},\mathcal{C}^k_{2i}, ... , \mathcal{C}^k_{Ni}]^T$, where $\mathcal{C}^k_{mi}$ represents the combined (coupled) gain of the $m$-th zone at time step $k$ for the $i$-th state. $\mathcal{C}^k_{mi}$ can be expressed as $\mathcal{C}^k_{mi} =  \{\xi_m(s_{i}^k) + \zeta_m(s_{i}^k)\}$, and ${J}^k_{i} = \mathcal{C}^{k=K}_{m(i=2n)}$ for the $m$-th zone. $\xi:S \rightarrow \mathbb{R}_{\ge 0}$ and $\zeta:S \rightarrow \mathbb{R}_{\ge 0}$ are functions that map each state $s_i^k \in S$ (from the DTMC of $m$-th zone ) as follows:  
		\[\xi_m(s_{i}^k)= 
		\begin{cases}
		\frac{Q_{int}}{C_m},& \text{if~} occupied \in L(s_{i}^{k-1})\\
		0,              & \text{otherwise}
		\end{cases}
		\]
		
		and,
		
		\[\zeta_m(s_{i}^k)= 
		\begin{cases}
		\frac{Q_{rad}}{C_m},& \text{if~} heating_{on} \in L(s_{i}^{k-1})\\
		0,              & \text{otherwise}
		\end{cases}
		\]
	\end{itemize}

	\noindent where $Q_{int}$  and $Q_{rad}$ are the per time step temperature gain of an occupied zone and from radiator from previous time step of a respective zone, and $C_m$ is the capacitance of the $m$-th zone. 
	
	

	
	
	
	

	
	
\end{definition}

\begin{remark} \label{rmk:heat}
	Note that, we use heating as a deterministic input to the model unlike the stochastic one as occupancy. In Section \ref{Sec:Res} we evaluate different heating strategies by providing the information if the heating is \emph{on} in that time step, which sets the value of $\zeta_m(s_i^k)$ for different time steps $k$ in the model for the $m$-th zone.
\end{remark}

\noindent Example 1. 
	Let us assume that we build the model for 9 time steps. So, using this model we can reason about the expected temperature at different time steps starting from time step 1 to time step 9. Consider a scenario where we are interested in reasoning about the expected temperature of the zone at time step $\theta=3$ and $\theta=8$ using the model that we built for 9 time steps. Since the thermal dynamics depends on the elapsed time, the state reward model needs to be dynamic as well. Hence, the parameter $\theta$ is used to assign the dynamic reward structure based on a specific time of interest. For the case $\theta=3$, the reward of state 0 is set to $\textbf{A}^3T[0]$, for states 1 and 2 the rewards are set to $\textbf{A}^2\mathcal{Q}^1_{1}$ and $\textbf{A}^2\mathcal{Q}^1_{2}$, for states 3 and 4 the rewards are set to $\textbf{A}\mathcal{Q}^2_{3}$ and $\textbf{A}\mathcal{Q}^2_{4}$, and finally, for state 5 and 6, the state rewards are set to $\mathcal{Q}^3_{5}$ and $\mathcal{Q}^3_{6}$. The reward for the rest of the states should be set to 0 since we are reasoning about the temperature at $\theta=3$, and hence we are interested in the cumulative reward from $k=0$ to $k=3$ while ignoring the rewards for other states. However, if we want to reason about the temperature of the same zone at time step $\theta=8$, the rewards for the same states will be set to different values. For this specific case the rewards from state 0 to state 4 need to be set as $\textbf{A}^8T[0]$, $\textbf{A}^7\mathcal{Q}^1_{1}$, $\textbf{A}^7\mathcal{Q}^1_{2}$, $\textbf{A}^6\mathcal{Q}^2_{3}$, $\textbf{A}^6\mathcal{Q}^2_{4}$ respectively, the rewards for state 5 to state 18 will also be assigned according to the Equation \eqref{eqn:reward}. Based on this example we can see that parameter $\theta$ can to be adjusted based on the time step of interest.

Given the DTMRM of each zone $\mathcal{R}_i \in \mathcal{B}$, the building model then can be constructed by using the parallel composition. Hence, if the building has $N$ rooms, the building model can be defined as $\mathcal{B} = \mathcal{R}_1 || \mathcal{R}_2 || ... || \mathcal{R}_N$.

\begin{definition}(Parallel Composition) \label{parr-comp}Given two DTMRM pairs  $\mathcal{R}_1 = (\mathcal{D}_1, \rho_1)$ and $\mathcal{R}_2 = (\mathcal{D}_2, \rho_2)$ with DTMC denoted as $\mathcal{D}_1 = (S_1, s_{init2}, TL_{1}, L_1, P_1)$ and  $\mathcal{D}_2 = (S_2, s_{init2},$  $ TL_{2}, L_2, P_2)$ with reward functions $\rho_1$ and $\rho_2$, their parallel composition is the system $\mathcal{R}_1 || \mathcal{R}_2  = (\tilde{S}, \tilde{s}_0, \tilde{T}L, \tilde{L}, \tilde{P})$ with reward function $\tilde{\rho}  $ where,
	
	\begin{itemize}
		\item $\tilde{S} = (S_1 \times  S_2);$
		\item $\tilde{s}_0 = (s_{init1}, s_{init2});$
		\item $\tilde{TL} = TL_{1} \cup TL_{2};$
		\item $\tilde{L} = L_1 \cup L_2;$
		
		\item \resizebox{0.7\columnwidth}{!}{$\tilde{P}=
		\begin{cases}
		\frac{s_1 \xrightarrow{\alpha, r} s_1' ~ \text{and}~ s_2 \xrightarrow{\alpha, q} s_2' }{(s_1,s_2) \xrightarrow{\alpha,rq} (s_1',s_2')}, & \mbox{if } TL_{1} = TL_{2} = \alpha \\ \\
		\frac{s_1 \xrightarrow{\alpha_1, r} s_1'}{(s_1,s_2) \xrightarrow{\alpha_1,r} (s_1',s_2)},
		\frac{s_2 \xrightarrow{\alpha_2, q} s_2'}{(s_1,s_2) \xrightarrow{\alpha_2,q} (s_1,s_2')}, & \mbox{otherwise} \\
		\end{cases}$}\\
		\item $\tilde{\rho}(s_i)$ = $\rho_1(s_i) \cup \rho_2(s_i)$  (refer to Remark \ref{rmk:reward}) 
	\end{itemize}
	where $s_1,s_1' \in S_1$, $\alpha_1 \in TL_{1}$, $\alpha_2 \in TL_{2}$, $\alpha \in TL_{1} \cap TL_{2}$, $r=p_1(s_1,s_1') $, $s_2$,$s_2' \in S_2$, $q=p_2(s_2,s_2')$.
\end{definition}

\begin{remark} \label{rmk:reward}
	Note that, we are interested in reasoning about the expected temperature of zones given that a reward function captures the temperature dynamics of a zone at a specific time step. Hence, each state in the composed model of a building $\mathcal{B}$ has $n$ separate reward functions given that the building has $N$ zones in it. This is unlike the approach introduced in the Markov reward automata \cite{guck2014modelling} where the rewards are added in the composed model. 
\end{remark}
 
As mentioned in section \ref{subs:RC}, zones in a building are connected via walls between them, and hence their thermal dynamics are also coupled. An alternative approach for modeling building dynamics with occupancy pattern will be to model both the zone explicitly into a single Markov chain (for the two zone case), instead of building two separate Markov chains and composing them together. However, the complexity of developing such models will increase dramatically with the increasing number of zones. Our proposed modeling approach is modular since each zone is modeled separately at first, and composed later. We use PRISM modeling language (the detail syntax of PRISM modeling language, and the detail of PRISM tool can be found in \cite{prism}) for modeling, and encode each zone of a building as separate \textit{modules}. This makes our model easily extendable to $N$ zones using the \textit{module renaming} feature in PRISM with minimal effort. The corresponding probabilistic model of the building with $N$ zones in it (composed DTMC in our case) is constructed as the parallel composition of \textit{modules} specified in PRISM language which is useful for the automated analysis of such models.

\subsection{Transition matrix for zone occupancy}
\begin{table*}[t]
	\centering
	\caption{Example of a time-dependent occupancy dataset}
	\label{dataset}
	\resizebox{0.75\textwidth}{!}{
	\begin{tabular}{|c|c|c|c|c|c|c|c|c|c|}
		\hline
		\multirow{2}{*}{\textbf{Day}} & \multicolumn{4}{c|}{\textbf{9am}}                          & \textbf{10am-4pm} & \multicolumn{4}{c|}{\textbf{5pm}}                          \\ \cline{2-10} 
		
		 & $v \text{~to~} f$     & $v\text{~to~}v$     & $f\text{~to~}v$     & $f\text{~to~}f$     & .....      &$v\text{~to~}f$     & $v\text{~to~}v$     & $f\text{~to~}v$     & $f\text{~to~}f$ \\ \hline
		1& \checkmark &            & NA           &  NA           & .....      & \checkmark &            &            &            \\ \hline
		2&            & \checkmark &NA            &  NA           & .....      & \checkmark &            &            &            \\ \hline
		3&            & \checkmark &   NA          & NA            & .....      &            & \checkmark &            &            \\ \hline
		4&            & \checkmark &   NA          & NA            & .....      &            & \checkmark &            &            \\ \hline
		5&            & \checkmark              & NA            & NA  & .....      &            &            & \checkmark &            \\ \hline
		6& \checkmark           &            &      NA                    & NA  & .....      &            &            & \checkmark &            \\ \hline
		7& \checkmark                  &            &    NA         & NA  &            & .....      &                        &            & \checkmark \\ \hline
		....                     & .....      & .....      & .....      & .....      & .....      & .....      & .....      & .....      & .....      \\ \hline
		....                     & .....      & .....      & .....      & .....      & .....      & .....      & .....      & .....      & .....      \\ \hline
		179&    \checkmark        &            & NA  &    NA         & .....      &            &            &            & \checkmark \\ \hline
		180&            &    \checkmark        &   NA          & NA  & .....      &            & \checkmark &            &            \\ \hline
	\end{tabular}
	}
\end{table*}

In this section, we show how the transition matrix $P$ of a zone can be derived from the occupancy dataset. The occupancy dataset of a zone should contain per-hour occupancy record of each zone in the building for a period of time (say, for the last 6 months). This dataset can be updated periodically with recent data to reflect any changes in the occupancy pattern. Before calculating the probability, it is required to process the dataset in a specific format as shown in Table \ref{dataset}, where we enlist (for each hour) how many times a zone was observed to be empty or occupied from either occupied or empty state for a period of time. In our case, the classes in the university rooms start from 9 am, so the rooms are unoccupied at 8 am which clarifies the NA (not applicable) entries in the table. Given that a zone is occupied at time step $k$, the probability that the zone will be either occupied or empty in time step $k+1$ can be calculated from the dataset as follows:
\begin{equation} \label{eq:prOcc}
P(\mathcal{Z}^{k+1} = \mathcal{X}|\mathcal{Z}^{k}  = f) =
\begin{cases}
\frac{\sum_{i=0}^\mathcal{W} f^k_{i} \hookrightarrow f_i^{k+1}}{\sum_{i=0}^\mathcal{W} {f_i^{k} \hookrightarrow f_i^{k+1} + \sum_{i=0}^\mathcal{W} f_i^{k} \hookrightarrow v_i^{k+1}}}, & \text{if~} \mathcal{X} = f\\
1- P(\mathcal{Z}^{k+1} = f|\mathcal{Z}^{k}  = f), & \text{if~} \mathcal{X} = v
\end{cases}
\end{equation}

\begin{figure*}[h!]
	\centering
	\begin{subfigure}[t]{0.35\textwidth}
		\centering
	{\resizebox{\textwidth}{!}{
%
%
\definecolor{mycolor1}{rgb}{0.00000,0.44700,0.74100}%
\definecolor{mycolor2}{rgb}{0.85000,0.32500,0.09800}%
\begin{tikzpicture}

\begin{axis}[%
width=\textwidth,
height=2.15in,
legend pos=south east,
scale only axis,
xmin=8,
xmax=17,
xlabel style={font=\color{white!15!black}},
xtick ={8,9,10,11,12,13,14,15,16,17},
xlabel={Time},
ymin=0,
ymax=1,
ylabel style={font=\color{white!15!black}},
ylabel={$P(\mathcal{R}^{k+1} = occupied | \mathcal{R}^{k} = empty)$},
axis background/.style={fill=white},
legend style={legend cell align=left, align=left, draw=white!15!black}
]

\addplot [color=mycolor2,line width=1.5pt]
  table[row sep=crcr]{%
8	0.00\\
9	0.40\\
10	0.50\\
11	0.50\\
12	0.30\\
13	0.60\\
14	0.40\\
15	0.30\\
16	0.70\\
17	0.40\\
};
\addlegendentry{Zone 1}

\addplot [color=mycolor1,dashed,line width=1.5pt]
  table[row sep=crcr]{%
8	0.00\\
9	0.8\\
10	0.7\\
11	0.6\\
12	0.1\\
13	0.1\\
14	0.75\\
15	0.30\\
16	0.30\\
17	0.20\\
};
\addlegendentry{Zone 2}

\end{axis}

\end{tikzpicture}%
}
	\vspace{-4mm}\caption{Occupied from empty.}\label{fig:temp1}}		
	\end{subfigure}
	\quad
	\begin{subfigure}[t]{0.35\textwidth}
		\centering
		{\resizebox{\textwidth}{!}{
%
%
\definecolor{mycolor1}{rgb}{0.00000,0.44700,0.74100}%
\definecolor{mycolor2}{rgb}{0.85000,0.32500,0.09800}%
\begin{tikzpicture}

\begin{axis}[%
width=\textwidth,
height=2.15in,
legend pos=south east,
scale only axis,
xmin=8,
xmax=17,
xtick ={8,9,10,11,12,13,14,15,16,17},
xlabel style={font=\color{white!15!black}},
xlabel={Time},
ymin=0,
ymax=1,
ylabel style={font=\color{white!15!black}},
ylabel={$P(\mathcal{R}^{k+1} = empty | \mathcal{R}^{k} = occupied)$},
axis background/.style={fill=white},
legend style={legend cell align=left, align=left, draw=white!15!black}
]

\addplot [color=mycolor2,line width=1.5pt]
  table[row sep=crcr]{%
8	1.00\\
9	0.60\\
10	0.43\\
11	0.48\\
12	0.70\\
13	0.50\\
14	0.70\\
15	0.80\\
16	0.80\\
17	0.70\\
};
\addlegendentry{Zone 1}

\addplot [color=mycolor1,dashed,line width=1.5pt]
  table[row sep=crcr]{%
8	1.00\\
9	0.20\\
10	0.85\\
11	0.72\\
12	0.10\\
13	0.10\\
14	0.50\\
15	0.50\\
16	0.39\\
17	0.07\\
};
\addlegendentry{Zone 2}

\end{axis}
\end{tikzpicture}%
}
		\vspace{-4mm}\caption{Occupied from occupied.}\label{fig:temp2}}
	\end{subfigure}
	\caption{Probability of zones being occupied.}\label{fig:occu}
	\vspace{-3mm}
\end{figure*}
\noindent where $\mathcal{W}$ represents the total number of days in the dataset, $\hookrightarrow$ represents a transition between occupied (denoted as `$f$') and empty zone (denoted as `$v$') on $i$-th day from time step $k$ to $k+1$. Thus, using Equation \eqref{eq:prOcc}, we obtain the value of $P_{{fv}}^{k}$ and $P_{{ff}}^{k}$ at time step $k$. Similarly, if the room is currently empty ($\mathcal{Z}_k = v$), then the probability that the room will be either occupied or empty in the next time step, can be calculated as follows:
\begin{equation} \label{eq:prOcc1}
P(\mathcal{Z}^{k+1} = \mathcal{X}|\mathcal{Z}^{k} = v) = 
\begin{cases}
\frac{\sum_{i=0}^\mathcal{W} v_i^{k} \hookrightarrow f_i^{k+1} }{\sum_{i=0}^\mathcal{W} {v_i^{k} \hookrightarrow f_i^{k+1}  + \sum_{i=0}^\mathcal{W} v_i^{k} \hookrightarrow v_i^{k+1}}}, & \text{if~} \mathcal{X} = f\\
1 - P(\mathcal{Z}^{k+1} = f|\mathcal{Z}^{k}  = v), & \text{if~} \mathcal{X} = v
\end{cases}
\end{equation}

We obtain the value of $P_{{vv}}^{k}$ and $P_{{vf}}^{k}$ at time step $k$ with the help of Equation \eqref{eq:prOcc1}. For instance, Table \ref{dataset} shows an example of such a table with hourly occupancy data from 9 am to 5 pm for 6 months ($\mathcal{W} =$180 days). Let us consider the first 7 days record from Table \ref{dataset}. Given that a zone is empty at 8 am and using Equations \eqref{eq:prOcc} and \eqref{eq:prOcc1}, we can derive the probability that the zone will be occupied at 9 am is 0.43 and the probability that the zone will be empty is 0.57.

\section{Quantitative Analysis} \label{Sec:Res}
In this section we present the quantitative results obtained from the two zone scenario as shown in Section 2. We use the PRISM probabilistic model checker for specification and analysis of our model. Since the developed PRISM model (using the PRISM modelling language) is parametric, it is possible to evaluate the model for different parameter values without any modification of the core model. After the parallel composition, our approach resulted into 132 states (for the two zone scenario). From our occupancy dataset, we evaluate the occupancy pattern as shown in Figure \ref{fig:occu} for zone 1 and zone 2 from 8 am to 5 pm. The initial temperature (at 8 am) of zone 1 and zone 2 are initialized to 18\degree C and 16\degree C respectively, the internal gain of a zone from occupants is assumed to be 0.7\degree C per hour and the gain from the radiator (e.g., heating is on) is assumed to be 1.5\degree C per hour. Both the zones are empty at 8 am and we evaluate the expected temperature of the zones from 9 am to 5 pm. In PRISM, such analysis can be performed by specifying a PCTL property for cumulative rewards as $\mathcal{\textbf{R}}_{=?}[\mathcal{\textbf{C}} \leq \Psi]$, where ($9\leq\theta\leq 17$) and $\Psi=\theta + 1$. Given that we intend to maintain the expected temperature of both zones between 20\degree C to 22\degree C, we evaluate six potential heating strategies as shown in Table \ref{tab:scn}. Strategies from S1 to S4 are self explanatory. In strategy S5, we turn on the radiator for heating in every alternating hours, e.g., the heating is on at 9 am, 11 am, 1 pm, and 3 pm. In contrast, in strategy S6 we turn on the heating only for two time steps, at 8 am (pre-heating the zones) and 9 am, and keep it off for the rest of time. As mentioned earlier, such heating strategies are preset based on the predefined thermal conditions. This falls under the framework of rule based control which is commonly employed in the heating of buildings~\cite{ALIMOHAMMADISAGVAND2018167}.

\begin{table}[t]
    \begin{minipage}{.5\linewidth}
    \centering
      \caption{Heating strategies}\label{tab:scn}
	\label{tab:scn}
	\resizebox{0.8\textwidth}{!}{
	\begin{tabular}{|c|c|c|}
		\hline
		\textbf{Strategies} & \textbf{\begin{tabular}[c]{@{}c@{}}Heating \\of zone 1\end{tabular}} & \textbf{\begin{tabular}[c]{@{}c@{}}Heating \\of zone 2\end{tabular}} \\ \hline
		S1                 & Off                                                                  & Off                                                                   \\ \hline
		S2                 & On                                                                   & On                                                                    \\ \hline
		S3                 & On                                                                   & Off                                                                   \\ \hline
		S4                 & Off                                                                  & On                                                                    \\ \hline
		S5                 & Alternating                                                          & Alternating                                                           \\ \hline
		S6                 & Selective                                                            & Selective                                                             \\ \hline
	\end{tabular}
	}
    \end{minipage}%
    \begin{minipage}{.5\linewidth}
    \centering
    \caption{Electricity tariff}\label{tab:plan}
        \resizebox{0.98\textwidth}{!}{
        \begin{tabular}{|c|c|c|}
		\hline
		\textbf{Time}              & \textbf{Plan}     & \textbf{Price/kW (\pounds)} \\ \hline
		8:00 am-9:59 am     & Economy  & 0.10          \\ \hline
		10:00 am - 12:59 pm & Off-peak & 0.15          \\ \hline
		1:00 pm - 5 pm      & Peak     & 0.20          \\ \hline
	\end{tabular}
	}
    \end{minipage} 
\end{table}

\begin{figure}[h!]
	\centering
	\begin{subfigure}[t]{.33\textwidth}
		\centering
		{\resizebox{\textwidth}{!}{
%
%
\definecolor{mycolor1}{rgb}{0.00000,0.44700,0.74100}%
\definecolor{mycolor2}{rgb}{0.85000,0.32500,0.09800}%
\begin{tikzpicture}

\begin{axis}[%
width=\textwidth,
height=2.15in,
at={(2.239in,0.602in)},
legend pos=south east,
scale only axis,
xmin=8,
xmax=17,
xlabel style={font=\color{white!15!black}},
xlabel={Time},
ymin=15,
ymax=20,
xtick ={8,9,10,11,12,13,14,15,16,17},
ylabel style={font=\color{white!15!black}},
ylabel={Expected temperature (\degree C)},
ytick ={12,13,14,15,16,17,18,19,20},
axis background/.style={fill=white},
legend style={legend cell align=left, align=left,draw=white!15!black},
legend pos=north west
]
\addplot [color=mycolor2,line width = 1.5pt]
  table[row sep=crcr]{%
8  18\\
9  17.400199999999998\\
10 17.364972\\
11 17.6457958772\\
12 18.002298797504\\
13 18.26200651711408\\
14 18.57091262969024\\
15 18.686886333122\\
16 18.898563030795724\\
17 19.294468909963662\\
};
\addlegendentry{Zone 1}

\addplot [color=mycolor1,dashed,line width=1.5pt]
  table[row sep=crcr]{%
8  16\\
9  16.6014\\
10 17.4016\\
11 17.964662660400002\\
12 18.357550089328\\
13 18.32077655976656\\
14 18.373037918724638\\
15 18.94016956915704\\
16 19.25236419650532\\
17 19.249059960463658\\
};
\addlegendentry{Zone 2}
\end{axis}
\end{tikzpicture}%
}\caption{Heating strategy S1.}\label{fig:tempS1}}
	\end{subfigure}~
	\begin{subfigure}[t]{.33\textwidth}
		\centering
		{\resizebox{\textwidth}{!}{
%
%
\definecolor{mycolor1}{rgb}{0.00000,0.44700,0.74100}%
\definecolor{mycolor2}{rgb}{0.85000,0.32500,0.09800}%
\begin{tikzpicture}

\begin{axis}[%
width=\textwidth,
height=2.15in,
at={(2.239in,0.602in)},
legend pos=south east,
scale only axis,
xmin=8,
xmax=17,
xlabel style={font=\color{white!15!black}},
xlabel={Time},
ymin=14,
ymax=32,
xtick ={8,9,10,11,12,13,14,15,16,17},
ylabel style={font=\color{white!15!black}},
ylabel={Expected temperature (\degree C)},
ytick ={12,14,16,18,20,22,24,26,28,30},
axis background/.style={fill=white},
legend style={legend cell align=left, align=left,draw=white!15!black},
legend pos=north west
]
\addplot [color=mycolor2,line width = 1.5pt]
  table[row sep=crcr]{%
8  18\\
9  17.400199999999998\\
10 18.459822000000003\\
11 19.9570003622\\
12 21.578394737504\\
13 23.12236008211408\\
14 24.723262184690242\\
15 26.134351533122\\
16 27.642359325795724\\
17 29.335082479963656\\
};
\addlegendentry{Zone 1}

\addplot [color=mycolor1,dashed,line width=1.5pt]
  table[row sep=crcr]{%
8  16\\
9  16.6014\\
10 18.901600000000002\\
11 20.8428340554\\
12 22.565234364328\\
13 23.838526514766556\\
14 25.19311518872464\\
15 27.05945449915704\\
16 28.669641336505318\\
17 29.69482353046366\\
};
\addlegendentry{Zone 2}
\end{axis}
\end{tikzpicture}%
}\caption{Heating strategy S2.}\label{fig:tempS2}}
	\end{subfigure}~
	\begin{subfigure}[t]{.33\textwidth}
		\centering
		{\resizebox{\textwidth}{!}{
%
%
\definecolor{mycolor1}{rgb}{0.00000,0.44700,0.74100}%
\definecolor{mycolor2}{rgb}{0.85000,0.32500,0.09800}%
\begin{tikzpicture}

\begin{axis}[%
width=\textwidth,
height=2.15in,
at={(2.239in,0.602in)},
legend pos=south east,
scale only axis,
xmin=8,
xmax=17,
xlabel style={font=\color{white!15!black}},
xtick ={8,9,10,11,12,13,14,15,16,17},
xlabel={Time},
ymin=14,
ymax=26,
ylabel style={font=\color{white!15!black}},
ytick ={14,16,18,20,22,24,26},
ylabel={Expected temperature (\degree C)},
axis background/.style={fill=white},
legend style={legend cell align=left, align=left,draw=white!15!black},
legend pos=north west
]
\addplot [color=mycolor2,line width=1.5pt]
  table[row sep=crcr]{%
8   18\\
9	17.400199999999998\\
10	18.459822000000003\\
11	19.5071503622\\
12	20.498994737504\\
13	21.34171008211408\\
14	22.21271218469024\\
15	22.882351533121998\\
16	23.644409325795724\\
17	24.58938247996366\\
};
\addlegendentry{Zone 1}

\addplot [color=mycolor1,dashed,line width=1.5pt]
  table[row sep=crcr]{%
8   16\\
9	16.6014\\
10	17.4016\\
11	18.293884055400003\\
12	19.147484364327997\\
13	19.62397651476656\\
14	20.21041518872464\\
15	21.32015449915704\\
16	22.17824133650532\\
17	23.44912353046366\\
};
\addlegendentry{Zone 2}
\end{axis}
\end{tikzpicture}%
}\caption{Heating strategy S3.}\label{fig:tempS3}}
	\end{subfigure}\\
	\begin{subfigure}[t]{.33\textwidth}
		\centering
		{\resizebox{\textwidth}{!}{
%
%
\definecolor{mycolor1}{rgb}{0.00000,0.44700,0.74100}%
\definecolor{mycolor2}{rgb}{0.85000,0.32500,0.09800}%
\begin{tikzpicture}

\begin{axis}[%
width=\textwidth,
height=2.15in,
at={(2.239in,0.602in)},
legend pos=south east,
scale only axis,
xmin=8,
xmax=17,
xtick ={8,9,10,11,12,13,14,15,16,17},
xlabel style={font=\color{white!15!black}},
xlabel={Time},
ymin=14,
ymax=27,
ylabel style={font=\color{white!15!black}},
ylabel={Expected temperature (\degree C)},
ytick ={14,16,18,20,22,24,26,28,30,32,34},
axis background/.style={fill=white},
legend style={legend cell align=left, align=left,draw=white!15!black},
legend pos=north west
]
\addplot [color=mycolor2,line width=1.5pt]
  table[row sep=crcr]{%
8   18\\
9	17.400199999999998\\
10	17.364972\\
11	18.095645877199996\\
12	19.081698797503996\\
13	20.04265651711408\\
14	21.081462629690243\\
15	21.938886333122003\\
16	22.89651303079572\\
17	24.04016890996366\\
};
\addlegendentry{Zone 1}

\addplot [color=mycolor1,dashed,line width=1.5pt]
  table[row sep=crcr]{%
8   16\\
9	16.6014\\
10	18.901600000000002\\
11	20.5136126604\\
12	21.775300089328002\\
13	22.53532655976656\\
14	23.35573791872464\\
15	24.67946956915704\\
16	25.74376419650532\\
17	25.494759960463657\\
};
\addlegendentry{Zone 2}
\end{axis}
\end{tikzpicture}%
}\caption{Heating strategy S4.}\label{fig:tempS4}}
	\end{subfigure}~
	\begin{subfigure}[t]{.33\textwidth}
		\centering
		{\resizebox{\textwidth}{!}{
%
%
\definecolor{mycolor1}{rgb}{0.00000,0.44700,0.74100}%
\definecolor{mycolor2}{rgb}{0.85000,0.32500,0.09800}%
\begin{tikzpicture}

\begin{axis}[%
width=\textwidth,
height=2.15in,
at={(2.239in,0.602in)},
legend pos=south east,
scale only axis,
xmin=8,
xmax=17,
xlabel style={font=\color{white!15!black}},
xtick ={8,9,10,11,12,13,14,15,16,17},
xlabel={Time},
ymin=14,
ymax=27,
ylabel style={font=\color{white!15!black}},
ylabel={Expected temperature (\degree C)},
ytick ={14,16,18,20,22,24,26,28},
axis background/.style={fill=white},
legend style={legend cell align=left, align=left,draw=white!15!black},
legend pos=north west
]
\addplot [color=mycolor2,line width=1.5pt]
  table[row sep=crcr]{%
8   18\\
9	17.400199999999998\\
10	18.459822000000003\\
11	18.8621503622\\
12	20.362040252503995\\
13	20.762618627114083\\
14	22.222650074690243\\
15	22.482614088122002\\
16	23.84663157079572\\
17	24.387013939963662\\
};
\addlegendentry{Zone 1}

\addplot [color=mycolor1,dashed,line width=1.5pt]
  table[row sep=crcr]{%
8   16\\
9	16.6014\\
10	18.901600000000002\\
11	19.3428340554\\
12	21.187062969327997\\
13	21.009013634766557\\
14	22.50487811372464\\
15	22.927614304157043\\
16	24.682196601505318\\
17	24.341604990463658\\
};
\addlegendentry{Zone 2}
\end{axis}
\end{tikzpicture}%
}\caption{Heating strategy S5.}\label{fig:tempS5}}
	\end{subfigure}~
	\begin{subfigure}[t]{.33\textwidth}
		\centering
		{\resizebox{\textwidth}{!}{
%
%
\definecolor{mycolor1}{rgb}{0.00000,0.44700,0.74100}%
\definecolor{mycolor2}{rgb}{0.85000,0.32500,0.09800}%
\begin{tikzpicture}

\begin{axis}[%
width=\textwidth,
height=2.15in,
at={(2.239in,0.602in)},
legend pos=south east,
scale only axis,
xmin=8,
xmax=17,
xlabel style={font=\color{white!15!black}},
xtick ={8,9,10,11,12,13,14,15,16,17},
xlabel={Time},
ymin=14,
ymax=26,
ylabel style={font=\color{white!15!black}},
ylabel={Expected temperature (\degree C)},
ytick ={14,16,18,20,22,24,26},
axis background/.style={fill=white},
legend style={legend cell align=left, align=left,draw=white!15!black},
legend pos=north west
]
\addplot [color=mycolor2,line width=1.5pt]
  table[row sep=crcr]{%
8   18\\
9	18.49505\\
10	19.676176485000006\\
11	20.127041817199995\\
12	20.551447877504\\
13	20.838260132114083\\
14	21.15802426469024\\
15	21.278333073122\\
16	21.491711400795722\\
17	21.88830603496366\\
};
\addlegendentry{Zone 1}

\addplot [color=mycolor1,dashed,line width=1.5pt]
  table[row sep=crcr]{%
8   16\\
9	18.1014\\
10	20.279771394999997\\
11	20.672346935400004\\
12	20.997128649328\\
13	20.933169554766557\\
14	20.97457289372464\\
15	21.537369439157036\\
16	21.847821921505318\\
17	21.842897085463658\\
};
\addlegendentry{Zone 2}
\end{axis}
\end{tikzpicture}%
}\caption{Heating strategy S6}\label{fig:tempS6}}
	\end{subfigure}
	\caption{Expected temperature from respective heating strategies.}\label{fig:HeatStrat}
\end{figure}

We plot the obtained results in Figure \ref{fig:HeatStrat}. In Figure \ref{fig:HeatStrat}(a) and Figure \ref{fig:HeatStrat}(b) we observe that if strategy S1 is followed, e.g., heating is off for both the zones, then the expected temperature of these zones gradually increases (due to the internal gain from occupancy), but stays below 20\degree C till 5 pm. In contrast, if S2 is followed, e.g., heating is turned on for both zones, then the temperature is expected to cross 22\degree C by 1 pm in zone 1, and by 12 pm in zone 2. Meanwhile, by the end of the day (5 pm), both zones are expected to reach approximately 30\degree C. While applying strategies S3 and S4 to our model as shown in Figure \ref{fig:HeatStrat}(c) and in Figure \ref{fig:HeatStrat}(d), we also observe that the expected temperature of both zones eventually crosses the maximum value of our intended threshold of 22\degree C, even though it never reaches 30\degree C when compared to the heating strategy S2. With  strategy S5 in Figure \ref{fig:HeatStrat}(e), the expected temperature of zone 1 and zone 2 crosses 22\degree C after 2 pm, whereas both the zones stay below 20\degree C before 12 pm. Therefore, all of these five strategies can cause discomfort to the occupants. Finally, we apply strategy S6 (we turn on heating only for two hours from 8 am) to our model the obtained result is shown in Figure \ref{fig:HeatStrat}(f). Interestingly, we observe that the temperature of both zones reaches almost 20\degree C by 9 am, and stay below 22\degree C till the end of the day. Hence, based on the comfort of the occupants, heating strategy 6 best fits our requirement. Similarly, any given heating strategy can be evaluated using the proposed approach without any major changes to the proposed model.\\

\noindent \textbf{Cost analysis:} We also evaluate these six heating strategies with respect to their cost. Let us consider the following electricity tariff as shown in Table \ref{tab:plan}, in which the hours of a day have been divided in economy, peak and off-peak hours based on the electricity demand. We only show the tariffs for 8 am to 5 pm based on our operating window. Economy tariff is the cheapest since the electricity demand is the lowest in that part of the day. In contrast, during peak hours the electricity demand is higher, hence the price is most expensive as well. Assuming that each zone has a 1 kW radiator installed for heating, the cost of the evaluated strategies are shown in Table \ref{tab:ccost}, where column 1 enlists the strategies, columns 2, 3, and 4 shows how much energy was consumed in total by both zones in those corresponding  hours (economy, peak, off-pick) and column 5 shows the total cost incurred by applying the associated heating strategies. The cost calculation process is straightforward: \emph{Total cost = kWs consumed in economy hours * price per kW in economy tariff + kWs consumed in off-peak hours * price per kW in off-peak tariff + kWs consumed in peak hours * price per kW in peak tariff}. According to the cost analysis, among the evaluated strategies S6 is not only the most appropriate one in terms of maintaining the comfort level, but also the most cost effective one. Compared to S2, S3, S4 and S5, strategy S6 is 13.5, 6.75, 6.75 and 6.5 times more cost efficient respectively.

\begin{table}[t]
\centering
\caption{Cost analysis for different strategies}
\label{tab:ccost}
\resizebox{0.6\textwidth}{!}{
\begin{tabular}{|c|c|c|c|c|}
\hline
\textbf{Strategies}& \multicolumn{3}{c|}{\textbf{Total electricity consumed}} & \textbf{Total } \\ \cline{2-4}
                            & \textbf{Economy}           & \textbf{Off-peak}           & \textbf{Peak}          &    \textbf{cost (\pounds)}                                \\ \hline
S1                          & NA               & NA                & NA           & 0                                 \\ \hline
S2                          & 2kW               & 6kW                & 8kW           & 2.70                              \\ \hline
S3                          & 1kW               & 3kW                & 4kW           & 1.35                              \\ \hline
S4                          & 1kW               & 3kW                & 4kW           & 1.35                              \\ \hline
S5                          & 2kW               & 4kW                & 4kW           & 1.30                              \\ \hline
S6                          & 2kW               & NA                & NA           & 0.20                              \\ \hline
\end{tabular}
}
\end{table}



\section{Discussion}
\vspace{-3mm}
\label{Sec:discuss}
As mentioned in the quantitative analysis section, our model for the two zone scenario requires 132 states. In contrast to our approach, the work in \cite{soudjani2015faust} requires the state-space of the two zones to be gridded over the possible temperature values the zone temperatures can be in. If we consider the temperature of the zones to range between 15 \degree C and 30 \degree C (reasonable temperature range the zones will be in) and assume a uniform gridding procedure with a maximal error of 0.5 \degree C, it will require $900$ states. Hence, we argue that our approach is more scalable in such cases. The proposed framework (as shown in \figurename ~\ref{fig:method}) is generic, which means $N$ number of zones can be modeled using the demonstrated approach, given that the total number of states does not cause a state space explosion of the PRISM tool. PRISM model checker includes multiple model checking engines, many of which are based on symbolic implementations (using binary decision diagrams and their extensions) which enables the probabilistic verification of models of up to $10^{10}$ states.

It is worth mentioning that for the sake of simplicity, in this paper we ignored the local weather prediction outside the building (heat exchange with outside environment through walls). This can be addressed by including the outside environment as a separate zone in our model, and thus modeling the heat exchange with the adjacent zones. Note that, our modeling can express both heating and cooling of zones, where the heating requires to be encoded as a positive reward, and cooling requires to be encoded as a negative reward in the DTMRM of a zone. The PRISM tool allows modeling and analysis of non-negative rewards/costs only, and hence there is no straightforward way to analyze the model with cooling using the PRISM tool. However, one way to overcome this limitation is to encode the cooling also as a positive reward, and calculate the cumulative reward until the target state is reached. Finally, the cumulative reward needs to be deducted from the initial temperature of a zone.  



\section{Conclusions}
\label{Sec:Conc}
In this paper we presented a framework for modeling and automated analysis of thermal dynamics of a building using the discrete-time Markov reward model formalism. The model is built using the stochastic occupancy pattern, building's thermal properties and the heating strategies to be evaluated. We evaluated six different heating strategies using the PRISM model checker and concluded that the ``selective strategy" is the best option for maintaining the occupants' comfort. Also, we performed a cost analysis that shows that the ``selective strategy" is the most cost-efficient option among the evaluated heating strategies. As a part of our future works, we plan to incorporate solar gain and heat exchange with outside environment in our model. We also aim to extend the framework for automated synthesis of heating strategies while minimizing the cost and maximizing the occupants' comfort.
%
%
%
%
\bibliographystyle{splncs04}
\bibliography{ref}
\end{document}